\documentclass[english,american,superscriptaddress,prd,a4paper,nofootinbib,11pt,eqsecnum,secnumarabic]{revtex4-2}
\usepackage{multirow,slashed,float,graphicx,caption,subcaption,hyperref,color}
\usepackage[dvipsnames]{xcolor}
\usepackage[utf8]{inputenc}
\usepackage{braket}
\hypersetup{colorlinks=true, linkcolor=BurntOrange, citecolor=BurntOrange,filecolor=BurntOrange, urlcolor=BurntOrange}
\usepackage{mathtools}

\newcommand{\be}{\begin{equation}}
\newcommand{\ee}{\end{equation}}
\newcommand{\bea}{\begin{eqnarray}}
\newcommand{\eea}{\end{eqnarray}}

\def\bsp#1\esp{\begin{split}#1\end{split}}

\newcommand{\matrixbb}[4]{\left(\hspace{-5 pt}\begin{tabular}{ c c } ${#1}$ & ${#2}$ \\ ${#3}$ & ${#4}$ \end{tabular}\hspace{-5 pt}\right)}

\usepackage{epsfig}
\usepackage{amsbsy}
\usepackage{varioref}
\usepackage{pifont}
\usepackage{amsmath,amssymb}
\usepackage{wrapfig}

\usepackage[normalem]{ulem}

\usepackage{mdframed}
\usepackage[export]{adjustbox}

\begin{document}

\title{Supersymmetry and Sum Rules in the Goldberger-Wise Model}

\author{R. Sekhar Chivukula}
\affiliation{UC San Diego, 9500 Gilman Drive,  La Jolla, CA 92023-0001, USA}

\author{Elizabeth H. Simmons}
\affiliation{UC San Diego, 9500 Gilman Drive,  La Jolla, CA 92023-0001, USA}

\author{Xing Wang}
\affiliation{UC San Diego, 9500 Gilman Drive,  La Jolla, CA 92023-0001, USA}

\begin{abstract}
In this work we demonstrate that the mixed gravitational and scalar sectors of the five-dimensional Goldberger-Wise (GW) model, in which the size of a warped extra dimension is dynamically determined, has a ``hidden" dual $N=2$ supersymmetric structure. This symmetry structure, a generalization of one found in the unstabilized Randall-Sundrum model, is a result of the spontaneously broken five-dimensional diffeomorphism invariance of the underlying gravitational theory. The supersymmetries relate the properties of the spin-1 and spin-0 modes ``eaten" by the massive spin-2 Kaluza-Klein states of the theory to the mode functions of the spin-2 modes. Because the symmetries relate the couplings and masses of the massive spin-2 states to those of the tower of physical spin-0 states of the GW model, they enable us to analytically prove the sum rule relations which ensure the tree-level scattering amplitudes of the massive spin-2 states will grow no faster than ${\cal O}(s)$. The analysis given here also explains the unconventional forms of the spin-0 mode equation, boundary condition(s), and normalization found in the GW model.
\end{abstract}


\maketitle
\flushbottom

\section{Introduction}

The possibility that there are extra-dimensions of space has been explored since the pioneering work of 
Kaluza and Klein \cite{Kaluza:1921tu,Klein:1926tv} almost a century ago.\footnote{For a review, see \cite{Appelquist:1987nr}.} Of particular interest recently has been possibility that extra-dimensional models might be relevant to solving the hierarchy problem \cite{Antoniadis:1990ew,ArkaniHamed:1998rs,Appelquist:2000nn}. In the context of the Randall-Sundrum (RS) model \cite{Randall:1999ee,Randall:1999vf}, the hierarchy problem is recast in terms of the warped geometry of a compact dimension. In particular, the extra dimension is taken to be an interval with boundaries (or ``branes") at each end; if the curvature and proper-length of the extra dimension are chosen correctly, the natural high-energy scale at a ``brane" located at one end of the extra-dimension can correspond to the Planck scale, while the scale at the other brane (due to the warped geometry) can be the TeV scale.

As originally formulated, however, the size of the extra-dimension in the RS model is arbitrary. While the RS model could incorporate a hierarchy, the size of the hierarchy was not determined by other physical parameters of the theory. Furthermore,  fluctuations of the five-dimensional metric in the RS model which correspond to (locally) changing the proper-length of the compact dimension gave rise to a phenomenologically unacceptable massless scalar particle in the four-dimensional effective theory, the radion.\footnote{In particular, in the RS model the radion has couplings suppressed by the TeV scale, and is experimentally excluded.}

Goldberger and Wise (GW) \cite{Goldberger:1999uk,Goldberger:1999un} introduced a model in which the size of the extra dimension was dynamically determined. In the GW model an additional five-dimensional real scalar field is added with potential energy terms on the branes and in the bulk. These potentials are chosen such that the background scalar-field value is different on the two branes, and hence the background scalar field has an expectation value which depends on position in the extra dimension. As shown by Goldberger and Wise, in this case the competition between the expectation values of kinetic and potential energy of this background field can naturally fix the proper-length of the extra dimension. This stabilization of the size of the extra dimension then implies that the mode corresponding to the radion becomes massive \cite{Goldberger:1999uk,Goldberger:1999un,Tanaka:2000er,Csaki:2000zn}.

In this paper we study the properties of the scalar and gravitational sectors of the GW model and the mode expansions which give rise to the massive spin-2 Kaluza-Klein (KK) gravitons and the tower of physical scalar states. We show that these sectors have a hidden dual $N=2$ supersymmetry (SUSY) structure which relates the mode expansions of the four-dimensional massive spin-2 fields to those of the spin-1 and spin-0 fields which are ``eaten" due to the spontaneous breaking of five-dimensional diffeomorphism invariance \cite{Dolan:1983aa}. This dual $N=2$ SUSY structure of the GW model is a generalization of one discovered by Lim, {\it et.~al.}~\cite{Lim:2007fy,Lim:2008hi} in the unstabilized RS model, which is in turn a generalization of a hidden $N=2$ SUSY structure uncovered in extra-dimensional gauge theory \cite{Lim:2005rc}.  

The existence of this dual supersymmetry in the GW model
depends crucially on the consistency between the eigenfunction equations of the modes and the equations of motion satisfied by the background metric and scalar fields. The symmetry analysis we present allows one to clearly separate the unphysical eaten scalar modes from the physical tower of GW scalars in the theory, and to understand the unconventional forms of the mode equation, boundary condition(s), and normalization of these modes \cite{Boos:2005dc,Boos:2012zz,Chivukula:2021xod,Chivukula:2022tla} in this model.\footnote{See also \cite{Kofman:2004tk} for a discussion of the identification of the radion in the RS model.}

Our findings have implications for the scattering amplitudes of the massive spin-2 Kaluza-Klein states in the GW model.
In general, the scattering of the helicity-zero massive spin-2 states grow as fast as ${\cal O}(s^5)$, where $s$ is the center of mass scattering energy-squared \cite{ArkaniHamed:2002sp,ArkaniHamed:2003vb,Hinterbichler:2011tt,deRham:2014zqa}.
As previously demonstrated \cite{Chivukula:2019rij,Chivukula:2019zkt,Chivukula:2020hvi,Foren:2020egq,Chivukula:2021xod,Chivukula:2022tla},\footnote{See also \cite{Bonifacio:2019ioc,Hang:2021fmp} for related work on the sum rules in unstabilized Ricci-flat extra-dimensional gravity models.} in a theory of compactified extra-dimensional gravity the masses and couplings of the massive spin-2 Kaluza-Klein states and the radion (in the RS model) or the tower of GW scalar states (in the GW model) satisfy a set of sum rule relations which enforce cancellations such that the full amplitude grows no faster
than ${\cal O}(s)$. While the sum rules have all been demonstrated numerically, the sum rules which have been confirmed analytically were those which could be re-written to depend only on the properties of the wavefunctions of the spin-2 modes. The single sum rule which depends explicitly on the coupling of the massive spin-2 states to the scalar states (the radion or GW tower) has, so far, resisted analytic proof.

The dual $N=2$ supersymmetry uncovered here relates the mode wavefunctions of the spin-2 states to those of the spin-1 and spin-0 states, thereby relating the couplings (which are computed in terms of overlap integrals of these mode functions) and masses of these particles. Using the results of the analysis here, we demonstrate that the couplings of the spin-2 KK modes with one another and with the radion or tower of GW scalar states, along with the masses of these spin-2 and spin-0 particles, obey the additional so-far unproven sum rule. Our results here complete the analytic demonstration that the scattering amplitudes of the massive spin-2 states in the RS and GW models grow no faster than ${\cal O}(s)$.

In the next section we review the dual $N=2$ SUSY structure discovered in ref. \cite{Lim:2007fy}. This allows us to set our notational conventions, introduce the symmetry structure in a simpler setting, and develop the machinery needed to prove the sum rules in the unstabilized RS model \cite{Chivukula:2019rij,Chivukula:2019zkt,Chivukula:2020hvi}. The third section shows how the dual $N=2$ SUSY structure generalizes to the GW model -- in particular, allowing us to separate the physical and unphysical states in the scalar sector. We show how 5D diffeomorphism invariance and gauge-fixing proceeds in the GW model in a manner entirely analogous to the RS model \cite{Lim:2007fy}. The fourth section describes how the supersymmetries, along with the completeness of the relevant mode expansions, can be used to prove the remaining scalar sum rule found in \cite{Chivukula:2019rij,Chivukula:2019zkt,Chivukula:2020hvi,Chivukula:2021xod,Chivukula:2022tla}, which had previously only been demonstrated numerically. The last section gives our conclusions, and we include appendices which connect our analysis more directly to that given in \cite{Boos:2005dc,Boos:2012zz,Chivukula:2019rij,Chivukula:2019zkt,Chivukula:2020hvi,Foren:2020egq,Chivukula:2021xod,Chivukula:2022tla}.

\section{SUSY in The Randall-Sundrum Model}

We first review the dual $N=2$ supersymmetric structure \cite{Lim:2007fy} of the spectrum of the unstabilized Randall-Sundrum (RS) model.

\subsection{Geometry, Field Definitions, and Quadratic Lagrangian}

 The 5D Lagrangian in RS1 model \cite{Randall:1999ee,Randall:1999vf} can be written as 
\begin{equation}
    \mathcal{L}_{\rm 5D}^{\rm (RS)} = \mathcal{L}_{EH} + \mathcal{L}_{\rm CC} + \Delta\mathcal{L}~,
\end{equation}
where ${\cal L}_{EH}$ is the five-dimensional Einstein-Hilbert Lagrangian, ${\cal L}_{CC}$ includes the bulk and brane cosmological constants, and $\Delta {\cal L}$ includes total derivative terms needed to create a well-defined variational principle for the action \cite{Chivukula:2020hvi,Dyer:2008hb}. We parametrize the RS metric in the conformal coordinates $(x^\mu,z)$ as 
\begin{equation}
    G_{MN} = e^{2A(z)}\begin{pmatrix}
        e^{-\kappa\hat{\varphi}/\sqrt{6}}(\eta_{\mu\nu}+\kappa\hat{h}_{\mu\nu}) & \frac{\kappa}{\sqrt{2}}\hat{A}_\mu \\
        \frac{\kappa}{\sqrt{2}}\hat{A}_\mu & -\left(1+\frac{\kappa}{\sqrt{6}}\hat{\varphi}\right)^2
    \end{pmatrix}~,
    \label{eq:background-metric}
\end{equation}
where the field $\hat{h}_{\mu\nu}(x^\alpha,z)$ is a four-dimensional spin-2 field, and $\hat{A}_{\mu}(x^\alpha,z)$ and $\hat{\varphi}(x^\alpha,z)$ are the four-dimensional spin-1 and spin-0 fields, respectively.
The extra-dimension is taken to be the interval $z_1 \le z \le z_2$, where we associate $z_1$ as the location of the ``Planck brane" and $z_2$ as the location of the ``TeV brane." Finally, the warp factor
\begin{equation}
   A(z) = -\ln(kz)~,
\end{equation}
satisfies the background geometry bulk Einstein equation
\begin{equation}
    A''-(A')^2=0~,
    \label{eq:RS1EinsteinEq}
\end{equation}
and the value of $\kappa$ is set by the bulk and brane cosmological constants.

Using these definitions,  we find the kinetic (quadratic) terms of the of the fluctuating fields in the metric can be written as
\begin{equation}
    S = \int d^4x~dz~e^{3A(z)}\left( \mathcal{L}_{h\mbox{-}h} + \mathcal{L}_{h\mbox{-}A} + \mathcal{L}_{h\mbox{-}\varphi} + \mathcal{L}_{A\mbox{-}A} + \mathcal{L}_{A\mbox{-}\varphi} + \mathcal{L}_{\varphi\mbox{-}\varphi} \right),
    \label{eq:RS1Lagrangian}
\end{equation}
with
\begin{alignat}{5}
    \mathcal{L}_{h\mbox{-}h} &=&~ \hat{h}_{\mu\nu}\left[\vphantom{\frac{1}{4}}\right.&\frac{1}{4}\left(\eta^{\mu\rho}\partial^\nu\partial^\sigma + \eta^{\mu\sigma}\partial^\nu\partial^\rho + \eta^{\nu\rho}\partial^\mu\partial^\sigma + \eta^{\nu\sigma}\partial^\mu\partial^\rho\right) \nonumber\\
    & & &-\frac{1}{2}\left(\eta^{\mu\nu}\partial^\rho\partial^\sigma + \eta^{\rho\sigma}\partial^\mu\partial^\nu\right) \nonumber\\
    & & & \left.- \frac{1}{4}\left(\eta^{\mu\rho}\eta^{\nu\sigma} + \eta^{\mu\sigma}\eta^{\nu\rho} - 2\eta^{\mu\nu}\eta^{\rho\sigma}\right)\left(\Box +D^\dagger D\right)\right]\hat{h}_{\rho\sigma},\label{eq:RS1hh}\\
    \mathcal{L}_{h\mbox{-}A} &=&~ \hat{h}_{\mu\nu}\left[\vphantom{\frac{1}{4}}\right.&\left.\frac{1}{\sqrt{2}}\left(\eta^{\mu\rho}\partial^\nu + \eta^{\nu\rho}\partial^\mu - 2\eta^{\mu\nu}\partial^\rho\right)D^\dagger\right]\hat{A}_\rho,\label{eq:RS1hA}\\
    \mathcal{L}_{A\mbox{-}A} &=&~ \hat{A}_{\mu}\left[\vphantom{\frac{1}{4}}\right.&\left.-\frac{1}{2}\left(\partial^\mu\partial^\nu-\eta^{\mu\nu}\Box\right)\right]\hat{A}_\nu,\label{eq:RS1AA}\\
    \mathcal{L}_{h\mbox{-}\varphi} &=&~ -\hat{h}_{\mu\nu}\left[\vphantom{\frac{1}{4}}\right.&\left.\sqrt{\frac{3}{2}}\eta^{\mu\nu}D^\dagger \overline{D}^\dagger \right]\hat{\varphi},\label{eq:RS1hphi}\\
    \mathcal{L}_{A\mbox{-}\varphi} &=&~ \hat{A}_{\mu}\left[\vphantom{\frac{1}{4}}\right.&\left.\sqrt{3}\ \partial^\mu\overline{D}^\dagger \vphantom{\frac{1}{4}}\right]\hat{\varphi},\label{eq:RS1Aphi}\\
    \mathcal{L}_{\varphi\mbox{-}\varphi} &=&~ \hat{\varphi}\left[\vphantom{\frac{1}{4}}\right.&\left.-\frac{1}{2}\Box +2 \overline{D} \overline{D}^\dagger \right]\hat{\varphi},\label{eq:RS1phiphi}
\end{alignat}

In the expressions above we have defined \cite{Lim:2007fy} the differential operators 
\begin{equation}
    D = \partial_z,\quad D^\dagger = -(\partial_z+3A'),\quad \overline{D} = \partial_z+A',\quad \overline{D}^\dagger=-(\partial_z+2A').
\end{equation}
With respect to the inner product implicit in the action of Eq.~(\ref{eq:RS1Lagrangian})
\begin{equation}
    \langle F(z) G(z) \rangle \equiv \int_{z_1}^{z_2} dz\, e^{3A(z)} F(z) G(z)~,
    \label{eq:innerproduct}
\end{equation}
these operators form (as implied by the notation) two Hermitian pairs
\begin{align}
    \langle F (DG) \rangle & = \langle (D^\dagger F) G \rangle~, \\
    \langle F  (\overline{D} G) \rangle & = \langle ({\overline D}^\dagger F) G \rangle~,
\end{align}
so long as $F$ and $G$ satisfy the boundary conditions
\begin{equation}
  F(z_2)G(z_2)-F(z_1)G(z_1)=0~.
\end{equation}

\subsection{Supersymmetric Structure and Mode Expansions}

It is reasonable, given the form of Eq.~(\ref{eq:RS1hh}), to consider expanding the field $\hat{h}_{\mu\nu}(x^\alpha,z)$ in terms of eigenmodes $f^{(n)}(z)$ of the operator $D^\dagger D$ on the interval $[z_1,z_2]$. Furthermore, given the form of Eq.~(\ref{eq:RS1hA}), we see that it would be convenient to expand the field $\hat{A}_\nu(x^\alpha,z)$ in terms of modes related to $Df^{(n)}(z)$. Motivated by $N=2$ supersymmetric quantum mechanics \cite{Witten:1981nf,Cooper:1994eh}, we consider the supersymmetric partner of the operator $D^\dagger D$. In particular, the operator $D D^\dagger$ will have  the {\it same} non-zero eigenvalues
\begin{align}
    D^\dagger D f^{(n)}=-(\partial_z+3A')\partial_z f^{(n)} &= m_n^2 f^{(n)}, \\
    D D^\dagger g^{(n)}=-\partial_z(\partial_z+3A') g^{(n)} & =  m_n^2 g^{(n)}, 
    \label{eq:fg-mode-equations}
\end{align}
where the SUSY structure implies that\footnote{The SUSY algebras are explicitly constructed in \cite{Lim:2007fy}.}
\begin{equation}
    \begin{cases}
        D f^{(n)} = m_ng^{(n)}~, \\
        D^\dagger g^{(n)} = m_nf^{(n)}~,
    \end{cases}
    \label{eq:susy-fg}
\end{equation}
relating the eigenfunctions of the two operators. 
We note, however, that the SUSY relations are empty for a zero-mode, which we will need to discuss separately.

Next, from the form of Eq.~(\ref{eq:RS1phiphi}), we see that we would like to expand the modes of $\hat{\varphi}(x^\alpha,z)$ in terms of the eigenmodes of the operator $\overline{D}\overline{D}^\dagger$. Remarkably, as noted by \cite{Lim:2007fy} and of crucial importance in uncovering the supersymmetry structure(s) of the mode equations for the fluctuating fields, the Einstein equation (\ref{eq:RS1EinsteinEq}) implies \cite{Lim:2007fy} that
\begin{equation}
    \overline{D}^\dagger \overline{D}-D D^\dagger =2(A')^2-2A''=0~.
    \label{eq:unstabilized-Dbar}
\end{equation}
Therefore the non-zero modes among the $g^{(n)}$ are automatically eigenmodes of $\overline{D}^\dagger \overline{D}$, the SUSY partner of the operator $\overline{D}\overline{D}^\dagger$,
and we can write
\begin{align}
    \overline{D}^\dagger \overline{D} g^{(n)}=  - (\partial_z+2A')(\partial_z+A')g^{(n)} & = m_n^2 g^{(n)}~,\\
    \overline{D}\overline{D}^\dagger k^{(n)}= -(\partial_z+A')(\partial_z+2A') k^{(n)} &= m_n^2 k^{(n)}, \label{eq:us-kmode-equation}
\end{align}
where the $k^{(n)}$ are the eigenmodes paired with $g^{(n)}$ through this {\it second} supersymmetric structure.
The corresponding SUSY relations for the non-zero modes are
\begin{equation}
    \begin{cases}
        \overline{D} g^{(n)} = m_nk^{(n)} \\
        \overline{D}^\dagger k^{(n)} = m_ng^{(n)}~.
    \end{cases}
    \label{eq:susy-gk}
\end{equation}

Next, we consider the boundary conditions that can be imposed. 
We must find boundary conditions such that the four operators $D^\dagger D$,$D D^\dagger$, $\overline{D}^\dagger \overline{D}$, and $\overline{D}\overline{D}^\dagger$ are Hermitian under the inner product defined in Eq.~(\ref{eq:innerproduct}). Consider first the mode equations for $f^{(n)}$ and $k^{(n)}$: the most general boundary conditions are of the form
\begin{align}
    B_i Df^{(n)}(z_i) + C_i f^{(n)}(z_i) & =  0~, \label{eq:boundaryi}\\
    B'_i \overline{D}^\dagger k^{(n)} + C'_i k^{(n)}(z_i) & =  0~,
\end{align}
for $i=1,2$. As noted in \cite{Lim:2007fy}, however, these conditions are in general not consistent with the SUSY relations in Eqs.~(\ref{eq:susy-fg}) and (\ref{eq:susy-gk}), which impose (for all non-zero eigenvalues) two separate and potentially conflicting conditions for $g^{(n)}(z_i)$:
\begin{align}
    B_i m^2_ng^{(n)}(z_i)+C_i D^\dagger g^{(n)}(z_i) & = 0~,\label{eq:BC1}\\
    B'_i m^2_n g^{(n)}(z_i)+C'_i \overline{D}g^{(n)}(z_i) & = 0\, \Rightarrow \nonumber \\
    (B'_im^2_n-2A'(z_i))g^{(n)}(z_i)+C'_i D^\dagger g^{(n)}(z_i) & = 0~, \label{eq:BC2}
\end{align}
where in the last line we have used the relation $\overline{D}=-D^\dagger - 2A'$.
Equations (\ref{eq:BC1}) and (\ref{eq:BC2}), which both specify boundary conditions for $g^{(n)}$,  are generally inconsistent given the differing dependence on $m^2_n$ and $A'(z_i)$. The unique boundary conditions that are consistent with SUSY \cite{Lim:2007fy} are then those with $C_i=C'_i=0$ and $B_i,\, B'_i \neq 0$, implying that  
\begin{equation}
    D f^{(n)} = g^{(n)} = \overline{D}^\dagger k^{(n)} = 0, \quad \text{at }z=z_1,z_2.
    \label{eq:rs-BCs}
\end{equation}

With these boundary conditions there exist massless mode for both $f^{(0)}$ and $k^{(0)}$, but not for $g^{(0)}$,
\begin{equation}
    f^{(0)}(z) = {\rm Const},\quad k^{(0)}(z) = \mathcal{N}e^{-2A(z)},\quad g^{(0)}(z) = 0~.
    \label{eq:massless-RS-states}
\end{equation}
As we will see, the mode associated with $f^{(0)}$ is the usual massless 4D graviton, while that associated with $k^{(0)}$ is the massless scalar radion of the RS model.
The infinite tower of non-zero modes, $f^{(n)}$, $g^{(n)}$, and $k^{(n)}$ for $n>0$ have, due to the SUSY relations, the same eigenvalues for their respective mode equations.

Inspired by this, and following \cite{Lim:2007fy}, we 
perform the following KK decomposition of the metric fluctuations
\begin{eqnarray}
    \hat{h}_{\mu\nu}(x^\alpha,z) =&& \sum\limits_{n=0}^{\infty}\hat{h}_{\mu\nu}^{(n)}(x^\alpha)f^{(n)}(z),\label{eq:KK_1u}\\
    \hat{A}_{\mu}(x^\alpha,z) =&& \sum\limits_{n=1}^{\infty}\hat{A}_{\mu}^{(n)}(x^\alpha)g^{(n)}(z),\label{eq:KK_2u}\\
    \hat{\varphi}(x^\alpha,z) =&&~ \hat{r}(x^\alpha)k^{(0)}(z) +  \sum\limits_{n=1}^{\infty}\hat{\pi}^{(n)}(x)k^{(n)}(z)~,\label{eq:KK_3u}
\end{eqnarray}
where we choose the normalizations of the modes $f^{(n)}(z)$, $g^{(n)}(z)$, and $k^{(n)}(z)$ with respect to the inner-product in Eq.~(\ref{eq:innerproduct}), $\langle f^{(n)} f^{(m)}\rangle = \delta_{n}$ and analogously for $g^{(n)}$ and $k^{(n)}$.
As we see below, the massive KK gravitons $\hat{h}^{(n)}_{\mu\nu}$ ($n>0$) acquire their masses by absorbing the KK Goldstone modes $\hat{A}^{(n)}_\mu$ and $\hat{\pi}^{(n)}$ with $n\neq 0$, as in the case when the internal manifold is toroidal or flat \cite{Dolan:1983aa,Bonifacio:2019ioc,Hang:2021fmp}.

\subsection{5D Diffeomorphism Invariance}

The 5D Lagrangian is invariant under an infinitesimal coordinate transformation,
\begin{equation}
    x^M\mapsto \overline{x}^M=x^M+\xi^M.
    \label{eq:linear-diffeomorphisms}
\end{equation}
At linearized level, the induced transformation on the metric and fields are
\begin{eqnarray}
     G_{MN}&\mapsto & G_{MN} - G_{MA}\partial_{N}\xi^{A} - G_{NA}\partial_{M}\xi^{A} - \xi^A\partial_AG_{MN}, \label{eq:RS-diff1} \\
     \hat{h}_{\mu\nu}&\mapsto & \hat{h}_{\mu\nu} - \partial_\mu\xi_\nu - \partial_\nu\xi_\mu - \eta_{\mu\nu}(\partial_z+3A')\xi^5,\\
     \hat{A}_{\mu}&\mapsto & \hat{A}_{\mu} - \sqrt{2}\partial_z\xi_\mu + \partial_\mu\xi^5,\\
     \hat{\varphi}&\mapsto & \hat{\varphi} - \sqrt{6}(\partial_z+A')\xi^5. \label{eq:RS-diff2}
\end{eqnarray}
Expanding the transformation parameters $\xi_\mu$ and $\xi^5$ using the eigenfunctions $f^{(n)}$ and $g^{(n)}$,
\begin{eqnarray}
     \xi_\mu(x^\alpha,z) &=& \sum\limits_{n=0}^{\infty}\xi^{(n)}_\mu(x^\alpha)f^{(n)}(z),\\
     \theta(x^\alpha,z) &\equiv& \xi^5(x^\alpha,z) = \sum\limits_{n=0}^{\infty}\theta^{(n)}(x^\alpha)g^{(n)}(z)~,
     \label{eq:diffeomorphism-modes}
\end{eqnarray}
the above transformations on the individual KK modes can be written as 
\begin{eqnarray}
     \hat{h}_{\mu\nu}^{(n)}&\mapsto & \hat{h}_{\mu\nu}^{(n)} - \partial_\mu\xi_\nu^{(n)} - \partial_\nu\xi_\mu^{(n)} + m_n \eta_{\mu\nu}\theta^{(n)},\label{eq:diffeomorphism-spin2}\\
     \hat{A}_{\mu}^{(n)}&\mapsto & \hat{A}_{\mu}^{(n)} - \sqrt{2}m_n\xi_\mu^{(n)} + \partial_\mu\theta^{(n)},\label{eq:diffeomorphism-spin1}\\
     \hat{\pi}^{(n)}&\mapsto & \hat{\pi}^{(n)} - \sqrt{6}m_n\theta^{(n)},\\
     \hat{r}&\mapsto & \hat{r}.
\end{eqnarray}
Note that $\hat{\pi}^{(n)}$ and $\hat{A}^{(n)}_\mu$ transform as Goldstone Bosons of the spontaneously broken 5D diffeomorphism transformations parameterized by $\xi^{(n)}_\mu$ and $\theta^{(n)}$ with $n>0$ \cite{Dolan:1983aa}.

Unitary gauge, in which the ``eaten" Goldstone fields $\hat{A}^{(n)}$ and $\hat{\pi}^{(n)}$ with $n\ge 1$ are set to zero, can be achieved by choosing
\begin{alignat}{2}
     \xi^{(n)}_\mu &=~ \dfrac{1}{\sqrt{2}m_n}\left(\hat{A}_\mu^{(n)} + \dfrac{1}{\sqrt{6}m_n}\partial_\mu\hat{\pi}^{(n)}\right),
     \quad &n\geq1,\label{eq:unitary1}\\
     \theta^{(n)} &=~ \dfrac{1}{\sqrt{6}m_n}\hat{\pi}^{(n)},\quad &n\geq 1, \label{eq:unitary2}
\end{alignat}
where these tranformations fix all five-dimensional diffeomorphisms modulo residual four-dimensional ones (which are generated by $\xi^{(0)}_\mu$). Correspondingly, we can redefine the KK graviton fields as
\begin{eqnarray}
     \tilde{h}_{\mu\nu}^{(n)} = 
        \hat{h}_{\mu\nu}^{(n)} - \dfrac{1}{\sqrt{2}m_n}\left(\partial_\mu\hat{A}_\nu^{(n)}+\partial_\nu\hat{A}_\mu^{(n)} + \sqrt{\dfrac{2}{3}}\dfrac{1}{m_n}\partial_\mu\partial_\nu\hat{\pi}^{(n)}\right) + \dfrac{1}{\sqrt{6}}\eta_{\mu\nu}\hat{\pi}^{(n)},\quad n\geq1. \label{eq:physical-spin2}
\end{eqnarray}
Since both $\tilde{h}_{\mu\nu}^{(n)}$ and $\hat{r}$ are invariant under these five-dimensional coordinate transformations (modulo four-dimensional diffeomorphisms), they are the physical degrees of freedom - the KK gravitons and the (massless) radion respectively. 

Alternatively, the gauge redundancy can be removed by introducing the 5D 't Hooft-Feynman gauge fixing term \cite{Lim:2008hi,Hang:2021fmp}
\begin{equation}
    \mathcal{L}_{\rm GF} = F_\mu F^\mu - F_5 F_5,
    \label{eq:tHooft-gauge}
\end{equation}
where
\begin{eqnarray}
    F_\mu &=& -\left(\partial^\nu \hat{h}_{\mu\nu} - \dfrac{1}{2}\partial_\mu \hat{h}^\nu_\nu + \dfrac{1}{\sqrt{2}}D^\dagger\hat{A}_\mu\right), \label{eq:Fmu} \\
    F_5 &=& -\left(\dfrac{1}{2}D \hat{h}^{\mu}_\mu - \dfrac{1}{\sqrt{2}}\partial_\mu \hat{A}^\mu +\sqrt{\dfrac{3}{2}}\overline{D}^\dagger\hat{\varphi}\right).
\end{eqnarray}
The gauge-fixed kinetic terms in 't-Hooft-Feynman gauge are then given by
\begin{equation}
\aligned
    S = \int d^4x~\sum_n\left\{\vphantom{\dfrac{1}{2}}\right.&\dfrac{1}{2}\hat{h}^{(n)}_{\mu\nu}\left[\dfrac{1}{2}\left(\eta^{\mu\rho}\eta^{\nu\sigma} + \eta^{\mu\sigma}\eta^{\nu\rho} - \eta^{\mu\nu}\eta^{\rho\sigma}\right)(-\Box - m_n^2)\right]\hat{h}^{(n)}_{\rho\sigma} \\
    & + \dfrac{1}{2}\hat{A}^{(n)}_{\mu}\left[-\eta^{\mu\nu}(-\Box - m_n^2)\right]\hat{A}^{(n)}_{\nu} \\
    & + \dfrac{1}{2}~\hat{\pi}\left(-\Box - m_n^2\right)\hat{\pi}\left.\vphantom{\dfrac{1}{2}}\right\} + \dfrac{1}{2}~\hat{r}\left(-\Box \right)\hat{r}.
\endaligned
\end{equation}
Note that as a consequence of the SUSY conditions all of the particles at each level $n$ are degenerate, as expected in 't-Hoft-Feynman gauge.

\section{SUSY in The Goldberger-Wise Model}

We now show that the dual $N=2$ SUSY structures uncovered in \cite{Lim:2007fy} for the (unstabilized) RS1 model generalize to the case in which the size of the extra dimension is dynamically stabilized via the Goldberger-Wise mechanism \cite{Goldberger:1999uk,Goldberger:1999un}.

\subsection{Fields and Quadratic Lagrangian}

The Goldberger-Wise mechanism \cite{Goldberger:1999uk,Goldberger:1999un} introduces a bulk scalar field $\hat{\Phi}$ with the kinetic term and potential terms
\begin{eqnarray}
    \mathcal{L}_{\Phi\Phi} &=& \sqrt{G}\left[\frac{1}{2} G^{MN} \partial_M\hat{\Phi}\partial_N\hat{\Phi}\right]~,\\
    \mathcal{L}_{\rm pot} &=& -\frac{4}{\kappa^2}\left[\sqrt{G}V[\hat{\Phi}] + \sqrt{\overline{G}}V_1[\hat{\Phi}]\delta_1(z-z_1)+ \sqrt{\overline{G}}V_2[\hat{\Phi}]\delta_1(z-z_2) \right]~.
\end{eqnarray}
The potential terms are chosen such that the ground state has a non-zero $z$-dependent expectation
value for $\hat{\Phi}$, and such that minimizing the action fixes the proper length of the extra dimension.
The bulk scalar field $\hat{\Phi}$ can be expanded around the background as 
\begin{equation}
    \hat{\Phi}(x^\alpha,z) = \frac{1}{\kappa} (\phi_0(z) + \hat{\phi}(x^\alpha,z)).
\end{equation}
The form of the background metric remains as in Eq.~(\ref{eq:background-metric}), and the background gravity and scalar field equations in the bulk $z_1<z<z_2$ are given by the Einstein equations and scalar equations of motion\footnote{The first equation, which is a convenient Einstein equation to use since it is independent of the bulk potential $V$, follows from the second two equations via the Bianchi identity.} 
\begin{eqnarray}
     &&A'^2-A'' = \dfrac{1}{12}(\phi'_0)^2~, \label{eq:bulk1}\\
     &&e^{2A}V = -6 A'^2+\dfrac{1}{8}(\phi'_0)^2~, \label{eq:bulk2}\\
     &&4e^{2A}\dot{V} = \phi_0''+3A'\phi_0'~, \label{eq:bulk3}
\end{eqnarray}
where $\dot{V}$ is the functional derivative $\delta V/\delta \phi$ evaluated at the background field configuration $\phi_0$.
As we discuss later, the presence of the brane potential terms lead to non-trivial boundary conditions for the scalar sector of the theory \cite{Tanaka:2000er,Csaki:2000zn,Kofman:2004tk,Boos:2005dc,Boos:2012zz}.
Note that these equations depend only on the {\it derivative} of the background scalar field
configuration, $\phi'_0$.

The quadratic terms of the fluctuating fields can be written as
\begin{equation}
    S = \int d^4x~dz~e^{3A(z)}\left( \mathcal{L}_{h\mbox{-}h} + \mathcal{L}_{h\mbox{-}A} + \mathcal{L}_{h\mbox{-}\phi/\varphi} + \mathcal{L}_{A\mbox{-}A} + \mathcal{L}_{A\mbox{-}\phi/\varphi} + \mathcal{L}_{\phi/\varphi\mbox{-}\phi/\varphi} \right),
\end{equation}
where, due to symmetry, terms involving only the spin-2 or spin-1 fields
only involve the warp factor $A$ and their form is unchanged from that given in Eqs.~(\ref{eq:RS1hh})-(\ref{eq:RS1AA}). In the presence of a non-trivial background with $\phi'_0\neq0$, there is mixing between the gravitational and scalar fields of the theory. The terms involving the combined scalar sector, the fields $\hat{\varphi}$ and $\hat{\phi}$ are 
\begin{alignat}{5}
    \mathcal{L}_{h\mbox{-}\phi/\varphi} &=&~ - \hat{h}_{\mu\nu}\left[\vphantom{\frac{1}{4}}\right.&\left.\sqrt{\frac{3}{2}}\eta^{\mu\nu}D^\dagger\left(\overline{\cal D}^\dagger \hat{\Psi}\right)_1\right],\\
    \mathcal{L}_{A\mbox{-}\phi/\varphi} &=&~ \hat{A}_{\mu}\left[\vphantom{\frac{1}{4}}\right.&\left.\sqrt{3}\ \partial^\mu\left(\overline{\cal D}^\dagger \hat{\Psi}\right)_1\right],\\
    \mathcal{L}_{\phi/\varphi\mbox{-}\phi/\varphi} &=&~ \hat{\Psi}\left[\vphantom{\frac{1}{4}}\right.&\left.-\frac{1}{2}\Box + \frac{1}{2}\overline{\cal D} \Lambda \overline{\cal D}^\dagger \right]\hat{\Psi}~,
    \label{eq:GWscalar-Lagrangian}
\end{alignat}
where  $D^\dagger =-(\partial_z+3A')$, as before. 
The linear operators $\overline{\cal D}$ and $\overline{\cal D}^\dagger$ are defined as 
\begin{equation}
   \overline{\cal D} = \begin{pmatrix}
        \partial_z + A' & -\dfrac{1}{\sqrt{6}}\phi_0'\\
        \dfrac{1}{\sqrt{6}}\phi_0' & -\dfrac{1}{\phi_0'}(\partial_z+2A')\phi_0'
    \end{pmatrix} , \quad
    \overline{\cal D}^\dagger = \begin{pmatrix}
        -(\partial_z + 2A') & \dfrac{1}{\sqrt{6}}\phi_0'\\
        -\dfrac{1}{\sqrt{6}}\phi_0' & \phi_0'(\partial_z+A')\dfrac{1}{\phi_0'}
        \end{pmatrix},
        \label{eq:calD-definition}
\end{equation}
acting on a two-component doublet
\begin{equation}
    \hat{\Psi}(x^\alpha,z) = \begin{pmatrix}
        \hat{\varphi}(x^\alpha,z)\\\hat{\phi}(x^\alpha,z)\end{pmatrix}.
\end{equation}
Here
\begin{equation}
    \Lambda = \begin{pmatrix}
        2& \\ & -1
    \end{pmatrix}~
\end{equation}
is a constant matrix, and
\begin{equation}
    \left(\overline{\cal D}^\dagger \hat{\Psi}\right)_1 = \left(-(\partial_z+2A')\hat{\varphi} + \frac{1}{\sqrt{6}}\phi_0'\hat{\phi}\right)~,
\end{equation}
is the upper component of the doublet $\overline{\cal D}^\dagger \hat{\Psi}$.

Note that when $\phi'_0\neq 0$ there is mixing
between  $\phi$ and $\varphi$, and that the operators $\overline{\cal D}$ and $\overline{\cal D}^\dagger $ are Hermitian conjugate to one another with respect to the two-component generalization of the inner product in Eq.~(\ref{eq:innerproduct}), 
\begin{equation}
    \langle \Upsilon \cdot  \overline{\cal D} \Xi\rangle = \int_{z_1}^{z_2} dz\, e^{3A(z)}\, \left[\Upsilon(z)\cdot \overline{\cal D} \Xi(z)\right] = \langle \overline{\cal D}^\dagger \Upsilon\cdot \Xi \rangle~,
    \label{eq:general-innerproduct}
\end{equation}
for two-component real vectors $\Upsilon(x^\alpha,z)$ and $\Xi(x^\alpha,z)$ such that 
\begin{equation}
    \Upsilon(z_2)\cdot \Xi(z_2)-\Upsilon(z_1)\cdot \Xi(z_1)=0~.
\end{equation}

\subsection{Supersymmetric Structure of the Goldberger-Wise Model}

Since the quadratic terms involving only the spin-2 or spin-1 fields are
the same as those in Eqs.~(\ref{eq:RS1hh})-(\ref{eq:RS1AA}), we will expand these
fields in terms of the eigenfunctions defined by Eq.~(\ref{eq:fg-mode-equations}), where the
operator is now defined in the warp-factor $A(z)$ satisfying Eq.~(\ref{eq:bulk1}). The scalar
sector of the GW model, however, is more complicated. 

Based on the form of the scalar-sector kinetic energy terms in Eq.~(\ref{eq:GWscalar-Lagrangian}), we should find the eigenfunctions and eigenvalues of  $\overline{\cal D} \Lambda \overline{\cal D}^\dagger$.  Inspired by the SUSY structure of the unstabilized model \cite{Lim:2007fy}, we will begin our analysis of the scalar sector by examining the operator  $ \overline{\cal D}^\dagger \overline{\cal D}$. We find that it is diagonal:
\begin{equation}
\aligned
    \overline{\cal D}^\dagger \overline{\cal D} &= \begin{pmatrix}
        -(\partial_z+2A')(\partial_z+A') + \dfrac{1}{6}\phi_0'^2 & \\
        & -\phi_0'(\partial_z+A')\dfrac{1}{\phi_0'^2}(\partial_z+2A')\phi_0' + \dfrac{1}{6}\phi_0'^2
    \end{pmatrix}\\
    &\equiv \begin{pmatrix}
        H&\\&\tilde{H}
    \end{pmatrix}.
\endaligned
\end{equation}
Furthermore, using the GW Einstein equation (\ref{eq:bulk1}), we find an immediate generalization
of Eq.~(\ref{eq:unstabilized-Dbar}) to the stabilized model
\begin{equation}
    DD^\dagger = -\partial_z(\partial_z+3A') = -(\partial_z+2A')(\partial_z+A') + \frac{1}{6}\phi_0'^2 = H.
    \label{eq:stabilized-Dbar}
\end{equation}
Hence we immediately see that the operator $H$ has the same non-zero eignemodes as the
operator $DD^\dagger$, and hence this portion of the scalar sector has the same non-zero eigenvalues as the spin-2 and spin-1 sector.

Since $ \overline{\cal D}^\dagger \overline{\cal D}$ is diagonal, the two-component eigenfunctions may be written as
\begin{equation}
    \left\{G^{(n)} = \begin{pmatrix}
        g^{(n)}\\0
    \end{pmatrix},
    \tilde{G}^{(n)} = \begin{pmatrix}
        0\\\tilde{g}^{(n)}
    \end{pmatrix}
    \right\},
\end{equation}
where $g^{(n)}$ and $\tilde{g}^{(n)}$ are the eigenfunctions of $H$ and $\tilde{H}$, respectively,
\begin{equation}
    \begin{cases}
        H g^{(n)} = m_n^2g^{(n)} \\
        \tilde{H} \tilde{g}^{(n)} = \tilde{m}_n^2\tilde{g}^{(n)},
    \end{cases} 
    \label{eq:susy_4}
\end{equation}
and, hence
\begin{align}
    \overline{\cal D}^\dagger \overline{\cal D} G^{(n)}=m^2_n G^{(n)}~,\\
    \overline{\cal D}^\dagger \overline{\cal D} \tilde{G}^{(n)}=\tilde{m}^2_n \tilde{G}^{(n)}~.
\end{align}
Note that, due to the relation in Eq.~(\ref{eq:stabilized-Dbar}),  the $g^{(n)}$'s are those given in the ``$f$-$g$" SUSY system of Eq.~(\ref{eq:susy-fg}) and $\tilde{g}^{(n)}$ are the additional
eigenfunctions of $\tilde{H}$ which will be needed to describe the GW scalar sector.

Based on the $N=2$ SUSY form of the operator $\overline{\cal D}^\dagger \overline{\cal D}$, we know
that the operator $\overline{\cal D} \overline{\cal D}^\dagger$ will have exactly the same eigenvalues (more precisely, only the non-zero eigenvalues are shared, but the GW scalar sector will contain only massive scalars). Hence we can construct two-component eigenvectors of $\overline{\cal D} \overline{\cal D}^\dagger$ 
\begin{align}
     \overline{\cal D}\overline{\cal D}^\dagger K^{(n)} &= m_n^2 K^{(n)}, \\
   \overline{\cal D}\overline{\cal D}^\dagger \tilde{K}^{(n)} &= \tilde{m}_n^2 \tilde{K}^{(n)}.
\end{align}
which can be computed directly from the eigenvectors $G^{(n)}$ and $\tilde{G}^{(n)}$ through
the SUSY relations which are the analogs of Eq.~(\ref{eq:susy-gk})
\begin{equation}
    \begin{cases}
        \overline{\cal D}G^{(n)} = m_nK^{(n)}~,\\
        \overline{\cal D}^\dagger K^{(n)} = m_nG^{(n)}~, \\
        \overline{\cal D} \tilde{G}^{(n)} = \tilde{m}_n\tilde{K}^{(n)}~, \\
        \overline{\cal D}^\dagger \tilde{K}^{(n)} = \tilde{m}_n\tilde{G}^{(n)}~.
    \end{cases}\label{eq:susy_6}
\end{equation}
Since the operators $\overline{\cal D}$ and $\overline{\cal D}^\dagger$ are Hermitian conjugate with respect to the generalized inner product in Eq.~(\ref{eq:general-innerproduct}), we will choose the vectors $K^{(n)}$ and $\tilde{K}^{(n)}$  to be normalized
\begin{align}
\langle K^{(n)} \cdot K^{(m)}\rangle &= \langle \tilde{K}^{(n)} \cdot \tilde{K}^{(m)}\rangle = \delta_{nm}\\
\langle K^{(n)}\cdot \tilde{K}^{(m)}\rangle & = 0~.
\end{align}

We write the eigenfunctions of $\overline{\cal D} \overline{\cal D}^\dagger$ in components as,
\begin{equation}
    K^{(n)}(z) = \begin{pmatrix}
        k^{(n)}(z)\\l^{(n)}(z)
    \end{pmatrix},\quad
    \tilde{K}^{(n)}(z) = \begin{pmatrix}
        \tilde{k}^{(n)}(z)\\\tilde{l}^{(n)}(z)
    \end{pmatrix},
\label{eq:K-components}
\end{equation}
In terms of these definitions we can explicitly compute the components of these modes in terms of the eigenfunctions $g^{(n)}$ and $\tilde{g}^{(n)}$. 
In particular, Eqs.~(\ref{eq:susy_6}) become
\begin{equation}
    \begin{dcases}
        -(\partial_z+2A') k^{(n)}+ \dfrac{1}{\sqrt{6}}\phi_0' l^{(n)} = m_ng^{(n)} \\
         -\dfrac{1}{\sqrt{6}}\phi_0' k^{(n)}+ \phi_0'(\partial_z+A') \dfrac{l^{(n)}}{\phi_0'}= 0 \\
       (\partial_z + A')g^{(n)} = m_n k^{(n)}\\
      \dfrac{1}{\sqrt{6}}\phi_0'g^{(n)} = m_n l^{(n)}
    \end{dcases},
    \label{eq:susy_7}
\end{equation}
\begin{equation}
    \begin{dcases}
        -(\partial_z+2A') \tilde{k}^{(n)}+ \dfrac{1}{\sqrt{6}}\phi_0' \tilde{l}^{(n)} = 0 \\
         -\dfrac{1}{\sqrt{6}}\phi_0' \tilde{k}^{(n)}+ \phi_0'(\partial_z+A') \dfrac{\tilde{l}^{(n)}}{\phi_0'}= \tilde{m}_n\tilde{g}^{(n)} \\
      -\dfrac{1}{\sqrt{6}}\phi_0'\tilde{g}^{(n)} = \tilde{m}_n \tilde{k}^{(n)} \\
      -\dfrac{1}{\phi_0'}(\partial_z + 2A')(\phi_0'\tilde{g}^{(n)}) = \tilde{m}_n \tilde{l}^{(n)}
    \end{dcases}.
    \label{eq:susy_8}
\end{equation}The functions $g^{(n)}$, which are associated with Eq.~(\ref{eq:fg-mode-equations}) are normalized as before, and we also normalize $\tilde{g}^{(n)}$ such that $\langle \tilde{g}^{(n)} \tilde{g}^{(m)}\rangle = \delta_{nm}$.

We then perform the following KK decomposition for the fields in the GW model
\begin{eqnarray}
    \hat{h}_{\mu\nu}(x^\alpha,z) =&& \sum\limits_{n=0}^{\infty}\hat{h}_{\mu\nu}^{(n)}(x^\alpha)f^{(n)}(z),\label{eq:KK_1}\\
    \hat{A}_{\mu}(x^\alpha,z) =&& \sum\limits_{n=1}^{\infty}\hat{A}_{\mu}^{(n)}(x^\alpha)g^{(n)}(z),\label{eq:KK_2}\\
    \hat{\Psi}(x^\alpha,z) =&& \sum\limits_{n=1}^{\infty}\hat{\pi}^{(n)}(x^\alpha)K^{(n)}(z) + \sum\limits_{n=0}^{\infty}\hat{r}^{(n)}(x)\tilde{K}^{(n)}(z),\label{eq:KK_3}
\end{eqnarray}
where the first two are precisely of the same form as in the Randall-Sundrum model, as shown
in Eqs.~(\ref{eq:KK_1u}) - (\ref{eq:KK_3u}) where the operators are defined in terms of the
GW model warp-factor $A(z)$.\footnote{Note that we have started the $\hat{A}^{(n)}$ and $\hat{\pi}^{(n)}$ sums at $n=1$ since, as we shall show, neither sector has a massless state. As we will also show, the lightest $\hat{r}^{(n)}$ will be massive as well, but is parametrically lighter than the other scalar states in the $\phi'_0\to 0$ limit -- in which case it corresponds to the massless radion $\hat{r}$ of the RS model.}

Computing $\overline{\cal D}^\dagger \hat{\Psi}$ and using the relations in Eq.~(\ref{eq:susy_6}),
we find
\begin{equation}
  \overline{\cal D}^\dagger \hat{\Psi}(x^\alpha,z) =   \begin{pmatrix}
    \sum\limits_{n=0}^{\infty}m_n \hat{\pi}^{(n)}(x^\alpha)g^{(n)}(z)\\
    \sum\limits_{n=0}^{\infty}\tilde{m}_n \hat{r}^{(n)}(x^\alpha)\tilde{g}^{(n)}(z)
    \end{pmatrix}~.
\end{equation}
Using the orthogonality of the mode functions $g^{(n)}$ and $\tilde{g}^{(n)}$, we find that
the terms $\hat{\Psi}\left[\overline{\cal D} \Lambda \overline{\cal D}^\dagger \right]\hat{\Psi}$ simplify
\begin{equation}
  \int d^4x dz\, e^{3A(z)}\,   \hat{\Psi}\left[\overline{\cal D} \Lambda \overline{\cal D}^\dagger \right]\hat{\Psi} = \sum_{n} \left[2 m^2_n (\hat{\pi}^{(n)})^2-\tilde{m}^2_n (\hat{r}^{(n)})^2\right]~.
\end{equation}
In total, the kinetic terms in the GW model involving the scalar
sector become
\begin{alignat}{5}
    \mathcal{L}^{\rm eff}_{h\mbox{-}\pi} &=&~ \sum_n \hat{h}^{(n)}_{\mu\nu}\left[\vphantom{\frac{1}{4}}\right.&\left.\sqrt{\frac{3}{2}}m_n^2\eta^{\mu\nu}\right]\hat{\pi}^{(n)},\\
    \mathcal{L}^{\rm eff}_{A\mbox{-}\pi} &=&~ \sum_n \hat{A}^{(n)}_{\mu}\left[\vphantom{\frac{1}{4}}\right.&\left.\sqrt{3}\ m_n\partial^\mu\right]\hat{\pi}^{(n)},\\
    \mathcal{L}^{\rm eff}_{\pi\mbox{-}\pi} &=&~ \sum_n \hat{\pi}^{(n)}\left[\vphantom{\frac{1}{4}}\right.&\left.-\frac{1}{2}(\Box - 2m_n^2)\right]\hat{\pi}^{(n)},\\
    \mathcal{L}^{\rm eff}_{r\mbox{-}r} &=&~ \sum_n \hat{r}^{(n)}\left[\vphantom{\frac{1}{4}}\right.&\left.-\frac{1}{2}(\Box + \tilde{m}_n^2)\right]\hat{r}^{(n)}.
    \label{eq:modeLag-rr}
\end{alignat}
Note that the scalar Goldstone bosons are given by the
fields $\hat{\pi}^{(n)}$ -- the $\hat{r}^{(n)}$ fields will be the physical scalar states
in the model.

\subsection{Boundary Conditions and SUSY}

Next we examine which boundary conditions respecting the Hermiticity of the operator $\overline{\cal D} \Lambda \overline{\cal D}^\dagger$, or equivalently $\overline{\cal D}\overline{\cal D}^\dagger$ by the arguments given above, are also consistent with the dual $N=2$ SUSY structures of the mode equations. We give the precise argument below, but the result is easy to anticipate since the modes associated with the Goldstone spin-0 modes are the two-component eigenfunctions $K^{(n)}$, hence these are partnered with spin-1 modes $g^{(n)}$ in one SUSY doublet, while the spin-1 modes $g^{(n)}$ are also partnered with the spin-2 modes $f^{(n)}$ through the second supersymmetry. Hence, the generaliztion of the consistent SUSY boundary conditions in the RS model, Eq.~(\ref{eq:rs-BCs}), will be
\begin{equation}
    D f^{(n)} = g^{(n)}\quad {\rm and} \quad  \overline{\cal D}^\dagger K^{(n)} =G^{(n)}=0, \quad \text{at }z=z_1,z_2.
    \label{eq:GW-BCs}
\end{equation}
The modes $\tilde{K}^{(n)}$ (and their SUSY partners $\tilde{G}^{(n)}$) on the other hand, are decoupled from
the gravitational sector in the quadratic terms of the theory (see Eq.~(\ref{eq:modeLag-rr})), and will be less constrained, as we now show.

The operator $\overline{\cal D}\overline{\cal D}^\dagger$ is Hermitian if
\begin{equation}
     \overline{\cal D}^\dagger \Psi(x^\alpha,z_i) + {\cal B}_i \Psi(x^\alpha, z_i) = 0, \quad {\rm at~} z_i = z_1, z_2~,
     \label{eq:bc_2}
\end{equation}
where ${\cal B}_i$ is an arbitrary $2 \times 2$ real matrix, and Dirichlet boundary conditions are 
obtained by taking entries of ${\cal B}$ to infinity.
In general, as discussed above in the Randall-Sundrum model, the boundary conditions $Df^{(n)}=g^{(n)}=0$ and those in (\ref{eq:bc_2}) are not compatible with the the supersymmetry relations given in Eqs.~(\ref{eq:susy-fg}) and (\ref{eq:susy_6}). 
Since these relations must hold for arbitrary fields $\hat{\pi}^{(n)}(x^\alpha)$ and $\hat{r}^{(n)}(x^\alpha)$, we can impose these conditions separately on the terms involving $K^{(n)}$ and $\tilde{K}^{(n)}$ in the mode expansion for $\hat{\Psi}$. Considering the terms proportional to $K^{(n)}$, and re-writing them in terms of $g^{(n)}$ using Eq.~(\ref{eq:susy_7}), we find
the relation
\begin{equation}
    \begin{pmatrix}
    m_ng^{(n)}(z_i)\\
    0
    \end{pmatrix} + {\cal B}_i
    \begin{pmatrix}
    \dfrac{\overline{D}g^{(n)}(z_i)}{m_n}\\
    \dfrac{1}{\sqrt{6}m_n} \phi'_0(z_i) g^{(n)}(z_i)
    \end{pmatrix}=0~.
\end{equation}
The argument now proceeds just as before: the combination of the boundary conditions for the spin-2/spin-1 system, Eq.~(\ref{eq:BC1}), which involves $D^\dagger g^{(n)}$, is incompatible
with the relation above unless $g^{(n)}(z_i)\equiv 0$. Imposing $g^{(n)}(z_i)=0$, we also see that the first column of ${\cal B}_i$ must also vanish to eliminate the nonzero contributions proportional to $\overline{D}g^{(n)}(z_i)$. 

Using the SUSY relations, the modes $g^{(n)}(z)$ can be derived either from Eq.~(\ref{eq:fg-mode-equations}) or (\ref{eq:susy_4}), and the boundary conditions $g^{(n)}(z_i)=0$. The SUSY conditions in Eqs.~(\ref{eq:susy_7}) then allow one to compute the individual components of $K^{(n)}$, which
satisfy the boundary conditions
\begin{equation}
      \begin{cases}\overline{D}^\dagger k^{(n)}(z_i) =0 \\
     l^{(n)}(z_i) = 0~.
     \end{cases} 
\end{equation}

Next, consider the terms in Eq.~(\ref{eq:bc_2}) proportional $\tilde{K}^{(n)}$. Using Eq.~(\ref{eq:susy_8}) we find
\begin{equation}
    \begin{pmatrix}
    0 \\
    \tilde{m}_n\tilde{g}^{(n)}(z_i)
    \end{pmatrix} + {\cal B}_i
    \begin{pmatrix}
    -\,\dfrac{1}{\sqrt{6}\tilde{m}_n} \phi'_0(z_i) \tilde{g}^{(n)}(z_i)\\
    \dfrac{\overline{D}^\dagger (\phi'_0 \tilde{g}^{(n)})(z_i)}{\tilde{m}_n\phi'_0(z_i)}
    \end{pmatrix}=0~.
    \label{eq:k-tilde-bc}
\end{equation}
From the argument above, we know that the first column of ${\cal B}_i$ is zero - now we see that the first row must be zero as well. There are no other constraints imposed by SUSY, however, and the fields associated with $\tilde{g}^{(n)}$ may have arbitrary Robin boundary conditions. The general form of ${\cal B}_i$ is hence given by
\begin{equation}
    {\cal B}_i = \begin{pmatrix}
       0&&0\\0&&\beta_i
   \end{pmatrix}.
\end{equation}

The modes $\tilde{g}^{(n)}(z)$ are determined by the eigenvalue equations for $\tilde{H}$ in Eq.~(\ref{eq:susy_4}), with the boundary conditions determined by Eq.~(\ref{eq:k-tilde-bc}) to be
\begin{equation}
\beta_i\, \overline{D}^\dagger (\phi'_0 \tilde{g}^{(n)})(z_i) = -\tilde{m}^2_n \phi'_0(z_i) \tilde{g}^{(n)}(z_i)~.
\label{eq:gtilde-bc}
\end{equation}
Using the SUSY relations (\ref{eq:susy_8}), the boundary conditions on the components of $\tilde{K}^{(n)}$ become
\begin{equation}
     \begin{cases}
        -\dfrac{1}{\sqrt{6}}\phi_0'(z_i)\tilde{k}^{(n)}(z_i) + \phi_0'(z_i)\overline{D}\,\dfrac{\tilde{l}^{(n)}(z_i)}{\phi_0'(z_i)} + \beta_i \tilde{l}^{(n)}(z_i) = 0 \\
        \overline{D}^\dagger \tilde{k}^{(n)}(z_i)+ \dfrac{1}{\sqrt{6}}\phi_0'(z_i) \tilde{l}^{(n)}(z_i) = 0~.
    \end{cases}
    \label{eq:ktilde-bcs}
\end{equation}

The values of the boundary condition coefficients $\beta_i$ depend on the brane potentials and the background field values. Translating the analysis in \cite{Boos:2005dc,Boos:2012zz,Chivukula:2021xod} into conformal coordinates and using our field definitions, we find
\begin{equation}
    \beta_i = \mp2e^{A(z_i)}\ddot{V}_i(\phi_0(z_i)) + \dfrac{\phi_0''(z_i)}{\phi_0'(z_i)}-A'(z_i),
\end{equation}
where $\ddot{V}_i$ are the second variational derivatives of the brane potentials
$\delta^2 V_i/\delta \phi^2$ evaluated at the background field value $\phi_0$. The $\beta_i \rightarrow \infty$ in the ``stiff wall limit," in which case the boundary conditions corresponding to $\tilde{g}^{(n)}$ are $\overline{D}^\dagger (\phi'_0 \tilde{g}^{(n)})(z_i)=0$, with $\tilde{l}^{(n)}(z_i)=\overline{D}^\dagger \tilde{k}^{(n)}(z_i)=0$. 

The analysis here demonstrates that, for the boundary conditions given above, the operator $\overline{\cal D} \Lambda \overline{\cal D}^\dagger$ is Hermitian. Hence the mode equations determining the properties of the scalar sector of
the GW model are Sturm-Liouville equations, and the completeness of these eigenstates (which is unclear in unitary
gauge \cite{Csaki:2000zn}) follows. Furthermore, since the boundary conditions found here are consistent with the dual $N=2$ SUSY structure of the GW model, they are necessarily diffeomorphism invariant as well -- as we examine in Sec. \ref{sec:GW-Diff}.

\subsection{Absence of Massless Scalar Modes}

We now show, using the SUSY relations, that the scalar sector of the GW model has no massless modes. If there were a massless spin-0 Goldstone boson $\hat{\pi}^{(0)}$, its wavefunction would be constrained by Eq.~(\ref{eq:susy_7}), which becomes for $m_0=0$
\begin{equation}
    \begin{dcases}
        -(\partial_z+2A') k^{(0)}(z)+ \dfrac{1}{\sqrt{6}}\phi_0'(z) l^{(0)}(z) = 0 \\
         -\dfrac{1}{\sqrt{6}}\phi_0'(z) k^{(0)}(z)+ \phi_0'(z)(\partial_z+A') \dfrac{l^{(0)}(z)}{\phi_0'(z)}= 0,
    \end{dcases}
    \label{eq:no-massless-state}
\end{equation}
or, equivalently, using Eq.~(\ref{eq:stabilized-Dbar})
\begin{equation}
    \partial_z(\partial_z+3A')\left(\dfrac{l^{(0)}(z)}{\phi_0'(z)}\right) = 0~.
    \label{eq:no-massless-state-2}
\end{equation}
This equation has no non-trivial solution compatible with the conditions $\overline{\cal D}^\dagger K^{(n)}(z_i)=G^{(0)}(z_i)=0$, which we saw in the last section is required to maintain the SUSY structure of the mode equations,  and which implies $l^{(0)}(z_i) = 0$. 
The lowest eigenstate in this sector is therefore related to $g^{(1)}$ and $h^{(1)}$, and is paired with the lowest mass spin-2 Kaluza-Klein state. 

Precisely the same considerations apply to the scalar $\hat{r}^{(0)}$.  First, we get relations exactly analogous to Eq.~(\ref{eq:no-massless-state}) for $\tilde{k}^{(n)}$, $\tilde{l}^{(n)}$, and $\tilde{g}^{(n)}$ if we assume $\tilde{m}_0=0$. Furthermore, using the second relation in Eq.~(\ref{eq:susy_8}) with $\tilde{m}_0=0$, the boundary condition in Eq.~(\ref{eq:ktilde-bcs}) implies $\tilde{l}^{(0)}(z_i)=0$.
Hence there are no massless states in this sector either. 
However, the states in this sector are not paired with any others through the SUSY relations.

The presence of the $\phi'_0$ in the denominator of Eq.~(\ref{eq:no-massless-state-2}) is reassuring,
as it implies that the relation to the RS model (in which there is a massless state) is nontrivial.
In particular, in the limit $\phi'_0 \to 0$,\footnote{More precisely, the second and third derivatives of $\phi_0$ must go to zero faster than $\phi'_0$  for the limit to exist.} 
\begin{equation}
    \tilde{H} = \phi_0'\overline{D}\dfrac{1}{\phi_0'^2}\overline{D}^\dagger\phi_0' + \dfrac{1}{6}\phi_0'^2 \to  \overline{D}\,\overline{D}^\dagger~,
\end{equation}
and the lightest state in the $\tilde{K}^{(n)}$ sector,  $\hat{r}^{(0)}$, becomes the radion ($\hat{r}$) of the Randall-Sundrum model described in Eq.~(\ref{eq:massless-RS-states}). Its mass will be parametrically smaller than those of the other GW scalars for small $\phi'_0$. As shown in \cite{Chivukula:2021xod}, in the $\phi'_0 \to 0$ limit the GW states $\hat{r}^{(n)}$ for $n\ge 1$ decouple the gravitational ``Goldstone" sector and couple as ordinary scalar fields.

To summarize, for non-zero $\phi'_0$ there is no massless spin-0 mode in the GW model -- as expected physically, since the size of the extra dimension is fixed by the dynamics.\footnote{The only massless mode in the GW model is the spin-2 graviton, corresponding to the mode $f^{(0)}$ with boundary condition $Df^{(0)}=0$.}

 \subsection{Diffeomorphism Invariance and Gauge-Fixing}
 
 \label{sec:GW-Diff}

The analysis of diffeomorphism invariance in the GW model follows closely along the discussion
given for the RS model above and in \cite{Lim:2007fy}. Starting from the transformations encoded in Eq.~(\ref{eq:linear-diffeomorphisms}), one expands the transformation parameters $\xi_\mu$ and $\xi^5$ using the eigenfunctions $f^{(n)}$ and $g^{(n)}$, into modes $\xi^{(n)}_\mu(x^\alpha)$ and $\theta^{(n)}(x^\alpha)$ precisely as in Eq.~(\ref{eq:diffeomorphism-modes}). The form of the linearized transformations
of the gravity sector is the same as in Eqs.~(\ref{eq:RS-diff1}) - (\ref{eq:RS-diff2}), augmented
by the transformations of the GW scalar field fluctuations
\begin{eqnarray}
     \hat{\phi}&\mapsto & \hat{\phi} - \xi^5\phi_0'.
\end{eqnarray}
 Under these transformations, the spin-2 and spin-1 sectors
transform as previously shown in Eqs.~(\ref{eq:diffeomorphism-spin2}) and (\ref{eq:diffeomorphism-spin1}), while the scalar sector becomes
\begin{eqnarray}
     \hat{\pi}^{(n)}&\mapsto & \hat{\pi}^{(n)} - \sqrt{6}m_n\theta^{(n)},\\
     \hat{r}^{(n)}&\mapsto & \hat{r}^{(n)}~,
\end{eqnarray}
hence we see that the mode expansion has correctly separated the spin-0 Goldstone states
$\hat{\pi}^{(n)}$ from the tower of physical scalars $\hat{r}^{(n)}$.
Unitary gauge can be achieved as before, per Eqs.~(\ref{eq:unitary1}) and (\ref{eq:unitary2}), and
we find the same diffeomorphism-invariant spin-2 state 
\begin{equation}
     \tilde{h}_{\mu\nu}^{(n)} = 
        \hat{h}_{\mu\nu}^{(n)} - \dfrac{1}{\sqrt{2}m_n}\left(\partial_\mu\hat{A}_\nu^{(n)}+\partial_\nu\hat{A}_\mu^{(n)} + \sqrt{\dfrac{2}{3}}\dfrac{1}{m_n}\partial_\mu\partial_\nu\hat{\pi}^{(n)}\right) + \dfrac{1}{\sqrt{6}}\eta_{\mu\nu}\hat{\pi}^{(n)},\quad n\geq1.
        \tag{\ref{eq:physical-spin2} revisited}
\end{equation}
Since both $\tilde{h}_{\mu\nu}^{(n)}$ and $\hat{r}^{(n)}$ are invariant under five-dimensional coordinate transformations, modulo four-dimensional diffeomorphisms, they are the physical degrees of freedom. Again, the massless gravtion $\hat{h}^{(0)}$ is not invariant under the diffeomorphisms $\xi^{(0)}_\mu$ which are the unbroken 4D diffeomorphisms.

5D 't Hooft-Feynman gauge can again be achieved using the gauge-fixing term in Eq.~(\ref{eq:tHooft-gauge})
\begin{equation}
    \mathcal{L}_{\rm GF} = F_\mu F^\mu - F_5 F_5,
\end{equation}
where $F_\mu$ has precisely the same form as Eq.~(\ref{eq:Fmu}) and
\begin{eqnarray}
    F_5 &=& -\left(\dfrac{1}{2}\partial_z \hat{h}^{\mu}_\mu - \dfrac{1}{\sqrt{2}}\partial_\mu \hat{A}^\mu +\sqrt{\dfrac{3}{2}}\left(\overline{\cal D}^\dagger \hat{\Psi}\right)_1\right).
\end{eqnarray}
In 't Hooft-Feynman gauge, the kinetic terms of the fields are given by
\begin{equation}
\aligned
    S = \int d^4x~\sum_n\left\{\vphantom{\dfrac{1}{2}}\right.&\dfrac{1}{2}\hat{h}^{(n)}_{\mu\nu}\left[\dfrac{1}{2}\left(\eta^{\mu\rho}\eta^{\nu\sigma} + \eta^{\mu\sigma}\eta^{\nu\rho} - \eta^{\mu\nu}\eta^{\rho\sigma}\right)(-\Box - m_n^2)\right]\hat{h}^{(n)}_{\rho\sigma} \\
    & + \dfrac{1}{2}\hat{A}^{(n)}_{\mu}\left[-\eta^{\mu\nu}(-\Box - m_n^2)\right]\hat{A}^{(n)}_{\nu} \\
    & + \dfrac{1}{2}~\hat{\pi}\left(-\Box - m_n^2\right)\hat{\pi} + \dfrac{1}{2}~\hat{r}^{(n)}\left(-\Box - \tilde{m}_n^2\right)\hat{r}^{(n)}\left.\vphantom{\dfrac{1}{2}}\right\},
\endaligned
\end{equation}
and include the entire tower of GW states $\hat{r}^{(n)}$.

\section{Radion and Scalar Coupling Sum Rules}

The hidden supersymmetry relations in the Randall-Sundrum and Goldberger-Wise models connect the properties of the scalar mode wavefunctions to the wavefunctions of the spin-2 and spin-1 modes. As we now show, these relationships allow us to derive sum rules relating the couplings of the physical scalar modes (the massless radion in the RS model, and the tower of scalars in the GW model). The relations derived are precisely those needed \cite{Chivukula:2019rij,Chivukula:2019zkt,Chivukula:2020hvi,Chivukula:2021xod} to show that the radion or GW-scalar couplings ensure both that all ${\cal O}(s^3)$ and ${\cal O}(s^2)$ growth in the scattering of helicity-0 massive spin-2 KK states cancels, and that the overall amplitude grows only as fast as ${\cal O}(s)$.

In the next subsection we briefly show how the couplings defined in our previous work are related to overlap integrals in terms of the mode wavefunctions used here. In the first section we describe the physical fields in unitary gauge, examine the normalization and completeness of the corresponding modes, and define the couplings between the states of the model needed to compute scattering amplitudes. The second subsection derives the radion sum rule in the (unstabilized) RS model. The final subsection shows how the computation generalizes in the GW model.

\subsection{Unitary-Gauge Fields and Mode Coupling Definitions}

In unitary gauge, $\hat{\pi}^{(n)}=\hat{A}^{(n)}_\mu=0$. The
physical states include the spin-2 fields in
\begin{equation}
    \hat{h}_{\mu\nu}(x^\alpha,z) = \sum\limits_{n=0}^{\infty}\hat{h}_{\mu\nu}^{(n)}(x^\alpha)f^{(n)}(z),\tag{\ref{eq:KK_1} revisited}
\end{equation}
in both the RS and GW models. In the case of the RS model, the only remaining scalar field in unitary gauge is the massless radion $\hat{r}(x)$, associated with wavefunction $k^{(0)}(z)$. 

In the GW model, in unitary gauge the scalar fields become
\begin{equation}
    \hat{\Psi}(x^\alpha,z) = \begin{pmatrix}
        \hat{\varphi}(x^\alpha,z)\\\hat{\phi}(x^\alpha,z)\end{pmatrix}=
        \hat{\Psi}(x,z) = \sum\limits_{n=0}^{\infty}\hat{r}^{(n)}(x)\tilde{K}^{(n)}(z)~.
\end{equation}
Since the components of $\tilde{K}^{(n)}$ satisfy the relations in Eq.~(\ref{eq:susy_8}), we
see that in unitary gauge the scalar fields satisfy the relation
\begin{equation}
    \left(\overline{\cal D}^\dagger \hat{\Psi}\right)_1=\overline{D}^\dagger \hat{\varphi}(x^\alpha,z)+\dfrac{1}{\sqrt{6}}\phi'_0(z) \hat{\phi}(x^\alpha,z) = 0~.
\end{equation}
This is (in conformal coordinates) equivalent to the gauge conditions imposed in the analyses in  \cite{Boos:2005dc,Boos:2012zz,Chivukula:2021xod}. Using this expression, we can eliminate the field $\hat{\phi}$
associated with the bulk scalar in terms of the gravitational sector scalar field $\hat{\varphi}$. In the case
of the fields $\hat{r}^{(n)}$ this allows us to rewrite the lower components of $\hat{\Psi}$ in terms of
the upper ones. In this case, we use the relation
\begin{equation}
    -(\partial_z+2A') \tilde{k}^{(n)}+ \dfrac{1}{\sqrt{6}}\phi_0' \tilde{l}^{(n)} = 0
\end{equation}
to eliminate $\tilde{l}^{(n)}$
\begin{equation}
    \tilde{l}^{(n)} = \dfrac{\sqrt{6}}{\phi'_0} (\partial_z+2A')\tilde{k}^{(n)}~,
    \label{eq:eliminate-ell}
\end{equation}
and associate the field $\hat{r}^{(n)}$ in unitary gauge entirely with the wavefunctions $\tilde{k}^{(n)}$.

In any Kaluza-Klein theory, the couplings between the 4D fields are proportional to overlap integrals of the corresponding mode functions. In computing the scattering amplitudes of massive spin-2 fields $\hat{h}_{\mu\nu}^{(n)}$ we are interested in their couplings with themselves, as well as the radion field $\hat{r}$ in the RS model
and the tower of scalars $\hat{r}^{(m)}$ in the GW model. Hence we must consider overlap integrals associated with the modes $f^{(n)}$ with the $k^{(0)}$ in the RS model, and with the modes $\tilde{k}^{(m)}$ in the GW model.

The normalization of the mode functions, and hence of all the relevant couplings, are fixed
by the requirement that they have canonically normalized kinetic-energy terms. For the spin-2 KK modes and the radion of the RS model, this  requirement is straightforward -- and follows directly from the normalizations previously imposed: $\langle f^{(n)} f^{(m)}\rangle = \delta_{nm}$ and $\langle (k^{(0)})^2\rangle =1$. The situation is different for the GW scalars and the mode function $\tilde{k}^{(n)}$, however, as here the normalization is based on the normalization condition
\begin{equation}
 \langle \tilde{K}^{(n)}\cdot  \tilde{K}^{(m)}\rangle =   \int_{z_1}^{z_2} dz\, e^{3A} \left(\tilde{k}^{(n)} \tilde{k}^{(m)}+\tilde{l}^{(n)} \tilde{l}^{(m)}\right)= \delta_{nm}~.
\end{equation}
Using Eq.~(\ref{eq:eliminate-ell}), we find the $\tilde{k}^{(n)}$ normalization conditions
\begin{equation}
    \int_{z_1}^{z_2} e^{3A}\left[\tilde{k}^{(n)} \tilde{k}^{(m)} +\dfrac{6}{(\phi'_0)^2} \left[(\partial_z+2A')\tilde{k}^{(n)}\right] \cdot \left[(\partial_z+2A')\tilde{k}^{(m)}\right]\right] = \delta_{nm}~.
\end{equation}
This is, in conformal coordinates, the form of the unconventional normalization conditions found necessary in \cite{Boos:2005dc,Boos:2012zz,Chivukula:2021xod}.

Due to the Lorentz invariance of the background metric in Eq.~(\ref{eq:background-metric}) at fixed $z$, the form of the self-couplings between the spin-2 modes or between these modes and the scalars 
can be written in a form that involves either two 4D space-time derivatives $\partial_\mu$ or two extra-dimensional derivatives $\partial_z$ acting on the spin-2 field. Translating the results of \cite{Chivukula:2019rij,Chivukula:2019zkt,Chivukula:2019zkt,Foren:2020egq,Chivukula:2021xod} to conformal coordinates, therefore, we find the spin-2 self-couplings are related to the following overlap integrals
 \begin{eqnarray}
     a_{ijk} &=& \braket{f^{(i)} f^{(j)} f^{(k)}} \label{eq:overlap_1}\\
     a_{ijkl} &=& \braket{f^{(i)} f^{(j)} f^{(k)} f^{(l)}} \label{eq:overlap_2}\\
     b_{ijk} &=& \braket{(\partial_z f^{(i)}) (\partial_zf^{(j)}) f^{(k)}} =m_i m_j \braket{g^{(i)} g^{(j)}f^{(k)}}\label{eq:overlap_3}
 \end{eqnarray}
where we have used Eq.~(\ref{eq:susy-gk}), and  with
 \begin{equation}
     \braket{\psi_1\psi_2\cdots}\equiv \int dz~e^{3A(z)}\psi_1(z)\psi_2(z)\cdots.
 \end{equation}
In addition, the scattering amplitudes of massive spin-2 particles involves couplings
between these particles with the scalars, and require the scalar-coupling overlap integrals
\begin{eqnarray}
     b_{ijr} &=& \braket{(\partial_z f^{(i)}) (\partial_zf^{(j)}) k^{(0)}}=m_im_j\braket{g^{(i)} g^{(j)}k^{(0)}} \label{eq:overlap_4}\\
     a_{m'n'(i)} &=& \braket{(\partial_z f^{(m)}) (\partial_zf^{(n)}) \tilde{k}^{(i)}}=m_im_j\braket{g^{(i)} g^{(j)}\tilde{k}^{(i)}}~. \label{eq:overlap_5}
 \end{eqnarray}
for the RS model (first line) and the GW model (second line).

Note that the wavefunctions $\{f^{(n)}\}$, $\{g^{(n)}\}$ form complete basis for functions $\psi(z)$ with corresponding boundary conditions at $z=z_1,z_2$,
\begin{eqnarray}
     &\psi(z') = \sum\limits_{n}f^{(n)}(z')\braket{ f^{(n)}\psi} \quad &{\rm if~}\partial_z\psi(z_i) = D\psi(z_i) = 0,\label{eq:complete_1}\\
     &\psi(z') = \sum\limits_{n}g^{(n)}(z')\braket{g^{(n)}\psi} \quad &{\rm if~}\psi(z_i) = 0.\label{eq:complete_2}
 \end{eqnarray}
In the RS model, the scalar wavefunctions $\{k^{(n)}\}$ also form a set of complete basis,
\begin{equation}
     \psi(z') = \sum\limits_{n}k^{(n)}(z')\braket{k^{(n)}\psi} \quad {\rm if~}(\partial_z+2A')\psi(z_i) =\overline{D}^\dagger\psi(z_i) = 0,\label{eq:complete_3}
\end{equation}
while in the GW model, the completeness of the scalar wavefunctions $\{K^{(n)},\tilde{K}^{(n)}\}$ reads,
\begin{equation}
     \Psi(z') = \sum\limits_{n} \left[K^{(n)}(z')\braket{K^{(n)}\cdot\Psi}+\tilde{K}^{(n)}(z')\braket{\tilde{K}^{(n)}\cdot\Psi}\right] \quad {\rm if~}\overline{D}^\dagger\Psi(z_i)+\mathcal{B}_i\Psi(z_i) = 0.\label{eq:complete_4}
 \end{equation}
Even though we work in unitary gauge, we will see that the completeness relations involving the Goldstone mode functions $k^{(n)}(z)$ in the RS model, and involving both the Goldstone mode functions $K^{(n)}(z)$ and the scalar mode functions $\tilde{K}^{(n)}(z)$ in the GW model, will be essential in deriving the sum rule relations we seek.

Using completeness and integration by parts ~\cite{Chivukula:2020hvi}, the couplings satisfy the following relations
 \begin{eqnarray}
     &&b_{nnj} = (m_n^2 - \dfrac{1}{2}m_j^2)a_{nnj},\label{eq:ab_1} \\
     &&\sum\limits_{j=0}^{\infty}a_{nnj}^2 = a_{nnnn},\label{eq:ab_2}\\
     &&\sum\limits_{j=0}^{\infty}a_{nnj}b_{nnj} = \dfrac{1}{3}m_n^2a_{nnnn}.\label{eq:ab_3}
 \end{eqnarray}
 From Eqs.~(\ref{eq:ab_1}-\ref{eq:ab_3}) and using completeness, one can show that
 \begin{eqnarray}
     m^4_n\braket{g^{(n)}g^{(n)}g^{(n)}g^{(n)}}
     &=&
          \sum\limits_{j=0}^{\infty}\left(m_n^2\braket{g^{(n)}g^{(n)}f^{(j)}}\right)^2 \\
          &=& \sum\limits_{j=0}^{\infty}b_{nnj}^2 = \dfrac{1}{4} \sum\limits_{j=0}^{\infty}m_j^4a_{nnj}^2 - \dfrac{1}{3}m_n^4a_{nnnn}
     \label{eq:sumrule_1}
 \end{eqnarray}

\subsection{Radion Sum Rules in the RS Model}

 It is known \cite{Chivukula:2019rij,Chivukula:2019zkt,Chivukula:2020hvi,Foren:2020egq} that the couplings in the RS model must satisfy certain sum rules to result in $\mathcal{O}(s)$ dependence of the scattering amplitudes of massive gravitons. In particular, in the case of elastic scattering of level-$n$ KK-gravitons ($nn \to nn$), the radion coupling to KK-gravitons ($b_{nnr}$) is related to the KK-graviton three- and four-point self-couplings ($a_{nnj}$, where $j$ refers to the KK-level of the intermediate state, and $a_{nnnn}$) by
 \begin{equation}
     \dfrac{5}{4}\sum\limits_{j=0}^{\infty} m_j^4 a_{nnj}^2 - \dfrac{4}{3}m_n^4 a_{nnnn} = 9 b_{nnr}^2 - m_n^4 a_{nn0}^2~.
     \label{eq:RS-sum rule}
 \end{equation}
  The above radion sum rule relates the couplings of the KK gravtions to that of the radion, and has been verified numerically~\cite{Chivukula:2020hvi}, but so far not proved analytically. In this subsection, we will prove the radion sum rule analytically, using the SUSY relations among the spin-2, spin-1, and spin-0 wavefunctions and their completeness.

 By combining the SUSY relations
 \begin{equation}
     \begin{dcases}
         -(\partial_z+3A')g^{(j)} = m_j f^{(j)} \\
         (\partial_z + A')g^{(j)} = m_j k^{(j)}
     \end{dcases},
     \label{eq:k-fg-relationi}
 \end{equation}
 one gets
 \begin{equation}
     k^{(j)} = -f^{(j)} -\dfrac{2A'}{m_j}g^{(j)}\quad{\rm for~}j\neq 0.
     \label{eq:k-fg-relation}
 \end{equation}
 Thus, for $j>0$,
 \begin{eqnarray}
     m_n^2\braket{g^{(n)}g^{(n)}k^{(j)}} &=& - m_n^2\braket{g^{(n)}g^{(n)}f^{(j)}} - \dfrac{2m_n^2}{m_j}\braket{A'g^{(n)}g^{(n)}g^{(j)}}~.
  \end{eqnarray}
 Since the wavefunctions $g^{(n)}$ vanish at the boundaries,
 the following surface integrals vanish
 \begin{equation}
     \int_{z_1}^{z_2}dz \,\left[\partial_z\left(e^{3A}f^{(n)}f^{(n)}g^{(j)}\right)\right] = \int_{z_1}^{z_2}dz\, \left[\partial_z\left(e^{3A}g^{(n)}g^{(n)}g^{(j)}\right)\right] = 0~.
 \end{equation}
 Using the SUSY relations in Eq.~(\ref{eq:susy-gk}), one gets
 \begin{eqnarray}
     \braket{A'g^{(n)}g^{(n)}g^{(j)}} = -\dfrac{m_j}{6}\left(\braket{f^{(n)}f^{(n)}f^{(j)}} + \braket{g^{(n)}g^{(n)}f^{(j)}}\right).
 \end{eqnarray}
 Hence
  \begin{eqnarray}
       m_n^2\braket{g^{(n)}g^{(n)}k^{(j)}} &=& -\dfrac{2m_n^2}{3}\braket{g^{(n)}g^{(n)}f^{(j)}} + \dfrac{m_n^2}{3}\braket{f^{(n)}f^{(n)}f^{(j)}}\\
     &=&-\frac{2}{3}b_{nnj} + \frac{m^2_n}{3}a_{nnj}\qquad({\rm for~}j> 0)~.
\label{eq:RS-new-relation}
 \end{eqnarray}

Note that Eq.  (\ref{eq:RS-new-relation}) involves the couplings of the spin-2 modes to the {\it unphysical}
Goldstone boson fields $\hat{\pi}^{(j)}$ in Eq.~(\ref{eq:KK_3u}). However, since $g^{(n)}(z_i)=0$, it follows that $\overline{D}^\dagger[g^{(n)}]^2(z_i)=0$. We can then use  $k^{(j)}$ completeness in combination with Eq.~(\ref{eq:RS-new-relation}), along with Eqs.~(\ref{eq:ab_1}-\ref{eq:ab_3}) to find
 \begin{eqnarray}
    m^4_n\braket{g^{(n)}g^{(n)}g^{(n)}g^{(n)}}&=& \sum\limits_{j=0}^{\infty}\left(m_n^2\braket{g^{(n)}g^{(n)}k^{(j)}}\right)^2\\
    &=& \left(m_n^2\braket{g^{(n)}g^{(n)}k^{(0)}}\right)^2 + \sum\limits_{j=1}^{\infty}\left(m_n^2\braket{g^{(n)}g^{(n)}k^{(j)}}\right)^2 \\
     &=& ~b_{nnr}^2 + \sum\limits_{j=1}^{\infty}\left(-\frac{2}{3}b_{nnj} + \frac{m^2_n}{3}a_{nnj}\right)^2 \\
     &=& ~b_{nnr}^2 - \left(-\frac{2}{3}b_{nn0} + \frac{m^2_n}{3}a_{nn0}\right)^2+\sum\limits_{j=0}^{\infty}\left(-\frac{2}{3}b_{nnj} + \frac{m^2_n}{3}a_{nnj}\right)^2 \\
     &=& ~b_{nnr}^2 + \dfrac{1}{9}\sum\limits_{j=0}^{\infty}m_j^4a_{nnj}^2 - \dfrac{5}{27}m_n^4a_{nnnn}-\dfrac{1}{9}m_n^4a_{nn0}^2.
     \label{eq:sumrule_2}
 \end{eqnarray}
 Equating Eqs.(\ref{eq:sumrule_1}) and (\ref{eq:sumrule_2}) one derives Eq.~(\ref{eq:RS-sum rule}).

\subsection{Radion Sum Rules in the GW Model}

In the case of the GW model, the radion sum rule of the RS model generalizes to \cite{Chivukula:2021xod,Chivukula:2022tla}
 \begin{equation}
     \dfrac{5}{4}\sum\limits_{j=0}^{\infty} m_j^4 a_{nnj}^2 - \dfrac{4}{3}m_n^4 a_{nnnn} = 9 \sum\limits_{i=0}^{\infty}a_{n'n'(i)}^2 - m_n^4 a_{nn0}^2.
     \label{eq:GW-sum rule}
 \end{equation}
 where the radion coupling in Eq.~(\ref{eq:RS-sum rule}) generalizes to a sum of the couplings-squared of the couplings of the GW scalars to the spin-2 fields, where $a$ and $b$ are defined by the overlap integrals in Eqs.~(\ref{eq:overlap_1})-(\ref{eq:overlap_3}) and (\ref{eq:overlap_5}). We show now that the $N=2$ SUSY structure we have uncovered in the GW model allows us to prove this sum rule in a manner analogous to the discussion in the RS model given above.

 In the case of the GW mechanism, the completeness relation for the scalar wavefunctions in Eq.~(\ref{eq:complete_4}) can be written
 \begin{equation}
     \Psi(z') = \sum\limits_{n}K^{(n)}(z')\braket{K^{(n)}\cdot\Psi}+\tilde{K}^{(n)}(z')\braket{\tilde{K}^{(n)}\cdot\Psi} \quad {\rm if~}\overline{\cal D}^\dagger\Psi+{\cal B}_i\Psi = 0 {\rm ~at~}z=z_1,z_2.
 \end{equation}
 Consider a wavefunction of the form $\Xi(z) = (\xi(z),0)^T$. To be consistent
 with the boundary conditions $\overline{\cal D}^\dagger \Xi (z_i)+ {\cal B}_i\Xi(z_i)=0$, the functions $\xi(z)$ must satisfy
 \begin{equation}
     \overline{D}^\dagger \xi(z_i)=\xi(z_i)=0~.
     \label{eq:special-BCs}
 \end{equation}
for $z_i=z_1,\, z_2$. For a function satisfying these constraints
 \begin{equation}
     \xi(z') = \sum\limits_{n=1}^{\infty}k^{(n)}(z')\braket{k^{(n)}\xi}+\sum\limits_{n=0}^{\infty}\tilde{k}^{(n)}(z')\braket{\tilde{k}^{(n)}\xi}~,
\label{eq:k-special-completeness}
 \end{equation}
 where $k^{(0)}$ has been excluded from the summation since the massless mode does not exist.
 
Noting that the function $\xi(z)=(g^{(n)}(z))^2$ satisfies both boundary conditions in Eq.~(\ref{eq:special-BCs}) since $g^{(n)}(z_i)=0$, we can use Eq.~(\ref{eq:k-special-completeness}) to write
\begin{eqnarray}
    m^4_n\braket{g^{(n)}g^{(n)}g^{(n)}g^{(n)}} &=& \sum\limits_{j=1}^{\infty}\left(m_n^2\braket{g^{(n)}g^{(n)}k^{(j)}}\right)^2 + \sum\limits_{j=0}^{\infty}\left(m_n^2\braket{g^{(n)}g^{(n)}\tilde{k}^{(j)}}\right)^2\\
     &=& ~\sum\limits_{j=0}^{\infty}\left(m_n^2\braket{g^{(n)}g^{(n)}{k}^{(j)}}\right)^2+\sum\limits_{j=1}^{\infty}a_{n'n'(j)}^2 ~.
 \end{eqnarray}
Furthermore, from  Eq.~(\ref{eq:susy_7}) we see that Eq.~(\ref{eq:k-fg-relation}) and hence Eq.~(\ref{eq:RS-new-relation}) also holds in the GW model. Hence we find a direct generalization of
Eq.~(\ref{eq:sumrule_2}).
\begin{eqnarray}
    m^4_n\braket{g^{(n)}g^{(n)}g^{(n)}g^{(n)}} 
     &=& ~\sum\limits_{j=0}^{\infty}a_{n'n'(j)}^2 + \dfrac{1}{9}\sum\limits_{j=0}^{\infty}m_j^4a_{nnj}^2 - \dfrac{5}{27}m_n^4a_{nnnn}-\dfrac{1}{9}m_n^4a_{nn0}^2~,
     \label{eq:sumrule_3}
 \end{eqnarray}
holds in the GW model.  Equating Eqs.(\ref{eq:sumrule_1}) and (\ref{eq:sumrule_3}) one derives Eq.~(\ref{eq:GW-sum rule}).

\section{Conclusions}

\label{sec:concl}

The analysis given here illustrates how the mixing of the scalar and gravitational sectors in the GW model generalizes the properties seen in gauge symmetry-breaking in the presence of multiple ``Higgs" fields. In the multiple-Higgs case, only one ``direction" in field space actually gets a vacuum expectation value (vev). This direction is projected onto different mass eigenstate Higgs fields, giving rise to the diverse couplings and phenomenology of such models; the ``eaten" Goldstone bosons are defined by looking at the directions connected to the vev via (broken) symmetry transformations. Similarly in the GW model, in which the position-dependent background scalar field  induces mixing between the bulk scalar field and modes in the five-dimensional metric, using the dual $N=2$ SUSY structure of the GW model discovered here lets us precisely identify which combinations correspond to the Goldstone modes ($\hat{\pi}^{(n)}$) and which corrsespond to physical scalar fields ($\hat{r}^{(n)}$). This ability to cleanly separate these modes allows us to understand the unconventional forms of the mode equations and normalization conditions found previously \cite{Boos:2005dc,Boos:2012zz,Chivukula:2021xod,Chivukula:2022tla}, and understand how the scalar-field boundary conditions are consistent with five-dimensional diffeomorphism invariance.

In this paper we have demonstrated that the mixed gravitational and scalar sectors of the five-dimensional Goldberger-Wise (GW) model, in which the size of a warped extra dimension is dynamically determined, has a ``hidden" dual $N=2$ SUSY structure. Generalizing the result found for the unstabilized Randall-Sundrum model \cite{Lim:2007fy}, we see that these symmetries are the result of the spontaneously broken five-dimensional diffeomorphism invariance of the underlying gravitational theory. The supersymmetries allow us to relate the couplings and masses of the massive spin-2 states to those of the radion of the RS model and the tower of physical spin-0 states of the GW model, and to analytically prove the heretofore unproven sum rule relation which must hold in order for the tree-level scattering amplitudes of the massive spin-2 states to grow no faster than ${\cal O}(s)$.

\section*{Acknowledgements}
RSC, EHS, and XW were supported, in part, by by the US National Science Foundation under Grant No. PHY-1915147.

\newpage
\appendix

\section{Comparing Coordinate and Field Definitions}

In this appendix, we explain how the results derived here, using conformal coordinates to uncover the $N=2$ SUSY structure of the RS and GW models, compares to that given in previous work \cite{Boos:2005dc,Boos:2012zz,Chivukula:2019rij,Chivukula:2019zkt,Chivukula:2020hvi,Foren:2020egq,Chivukula:2021xod} using a non-conformal coordinate system. 

\subsection{Coordinates}

In our previous work \cite{Chivukula:2019rij,Chivukula:2019zkt,Chivukula:2020hvi,Foren:2020egq,Chivukula:2021xod} we analyzed warped extra-dimensional theories using coordinates $(x^\mu,y)$ corresponding to the metric
\begin{equation}
    ds^2 = e^{-2\tilde{A}(y)}\eta_{\mu\nu} dx^\mu dx^\nu - dy^2~,
\end{equation}
where $\tilde{A}(y)=ky$ in the case of (unstabilized) RS1, with $0\le y\le \pi r_c$.\footnote{More precisely, we considered the interval $-\pi r_c \le y \le \pi r_c$ with an orbifold identification $y\equiv -y$.} In this paper, we use conformal coordinates
\begin{equation}
    ds^2 = e^{+2A(z)}\left(\eta_{\mu\nu} dx^\mu dx^\nu-dz^2\right)~,
\end{equation}
as defined in \cite{Lim:2007fy}. Here, for RS1, 
\begin{equation}
    A(z)=-\ln\left(kz\right)\, \Rightarrow \, e^{2A(z)}=\left(\dfrac{1}{kz}\right)^2~=\left(\dfrac{z_1}{z}\right)^2,
\end{equation}
on the interval $z_1 \le z \le z_2$. The coordinate transformation between these systems is given by
\begin{equation}
    kz = e^{ky}~,
\end{equation}
with, therefore, $z_1=1/k$ and $z_2=e^{k\pi r_c}/k$.

\subsection{Fields and Mode Equations}

\subsubsection{Spin-2 Fields}

In this paper, we have
\begin{equation}
    G_{MN} = e^{2A(z)} \begin{pmatrix}
        e^{-\kappa\hat{\varphi}/\sqrt{6}}(\eta_{\mu\nu}+\kappa\hat{h}_{\mu\nu}) & \frac{1}{\sqrt{2}}\hat{A}_\mu \\
        \frac{1}{\sqrt{2}}\hat{A}_\mu & -\left(1+\frac{1}{\sqrt{6}}\hat{\varphi}\right)^2
    \end{pmatrix},
\end{equation}
for the coordinates $(x^\mu,z)$ while in our previous work we used
\begin{align}
    [G_{MN}] = \matrixbb{w\,g_{\mu\nu}}{0}{0}{-v^{2}}
    \label{eq:Metric}
\end{align}
where (taking our parameterization from \cite{Charmousis:1999rg}),
\begin{align}
     w = e^{-2[\tilde{A}(y)+\hat{u}(x,y)]}~,\hspace{35 pt}&v = 1 + 2\hat{u}(x,y)~,\\  g_{\mu\nu} = \eta_{\mu\nu} + \kappa\, \hat{h}_{\mu\nu}(x,y)~,\hspace{35 pt}&\hat{u} = \dfrac{ e^{2\tilde{A}(y)}}{2\sqrt{6}} \kappa\,\hat{r}(x,y)~. \label{eq:uhat}
\end{align}
for coordinates $(x^\mu,y)$. In the weak field, approximation, therefore, we see that
\begin{equation}
    \hat{h}_{\mu\nu}(x,z) \rightarrow \hat{h}_{\mu\nu}(x,y)~,
\end{equation}
and we compare the spin-2 mode eigenfunctions directly. 
Hence, the spin-2 equation described here
\begin{align}
    (\partial_z+3A')\partial_z f^{(n)} = e^{-3A} \partial_z \left(e^{3A} \partial_z f^{(n)}\right)=-m^2_n f^{(n)}~,
\end{align}
becomes, using the coordinate transformations above,
\begin{equation}
    \partial_y(e^{-4\tilde{A}}\partial_y f^{(n)})=-m^2_n e^{-2\tilde{A}}f^{(n)}~,
\end{equation}
in agreement with previous analyses.

\subsubsection{Spin-0 Sector}

In the weak-field approximation, comparing the definitions of the metric in the two coordinate system, we see the relation between the spin-0 metric fields are
\begin{equation}
    \hat{\varphi}(x,z) \rightarrow e^{2\tilde{A}(y)}\hat{r}(x,y)~,
\end{equation}
where we in \cite{Chivukula:2019rij,Chivukula:2019zkt,Chivukula:2020hvi,Foren:2020egq,Chivukula:2021xod} we defined the mode wavfunctions $\gamma^{(n)}(y)$ to expand $\hat{r}(x,y)$.
The mode wavefunctions for the physical scalar component in $\hat{\varphi}(z)$ in the current conformal-coordinate analysis are given by $\tilde{k}^{(n)}(z)$, however the mode-equation in conformal coordinates is most easily written in terms of $\tilde{g}^{(n)}(z)$, where
\begin{equation}
    \left(-\phi_0'(\partial_z+A')\dfrac{1}{\phi_0'^2}(\partial_z+2A')\phi_0' + \dfrac{1}{6}\phi_0'^2\right)\tilde{g}^{(n)} = \tilde{m}^2_n \tilde{g}^{(n)}~.
    \label{eq:gtilde-mode-equation}
\end{equation}
Furthermore, from
\begin{equation}
     -\dfrac{1}{\sqrt{6}}\phi_0'\tilde{g}^{(n)} = \tilde{m}_n \tilde{k}^{(n)}~,
\end{equation}
hence (recalling that $A(z) \leftrightarrow -\tilde{A}(y)$) we find the relationship between the two mode functions is given by 
\begin{equation}
\tilde{m}_n \gamma^{(n)}(y) \to  -\dfrac{1}{\sqrt{6}} \phi'_0(z) e^{2A(z)}\tilde{g}^{(n)}(z)~.
\label{eq:gammatogn}
\end{equation}
Rewriting Eq.~(\ref{eq:gtilde-mode-equation}) as
\begin{equation}
    e^{-A}\partial_z\left[\dfrac{e^{-A}}{(\phi'_0)^2}\partial_z \left(e^{2A}\phi'_0\tilde{g}^{(n)}\right)\right]-\dfrac{e^{-2A}}{6}\left(e^{2A}\phi'_0\tilde{g}^{(n)}\right)=-\tilde{m}^2_n \,\dfrac{e^{-2A}}{(\phi'_0)^2}\left(e^{2A}\phi'_0\tilde{g}^{(n)}\right)~,
\end{equation}
and applying our change of coordinates (remembering that $\phi'_0(z) = e^{-\tilde{A}(y)}\partial_y \phi_0$), we find the mode equation becomes rather unconventional
\begin{equation}
    \partial_y\left[\dfrac{e^{2\tilde{A}}}{(\partial_y\phi_0)^2}\partial_y\gamma^{(n)}\right]-\dfrac{e^{2\tilde{A}}}{6}\gamma^{(n)}=-\tilde{m}^2_n \dfrac{e^{4\tilde{A}}}{(\partial_y\phi_0)^2} \gamma^{(n)}~,
\end{equation}
in agreement with previous results \cite{Boos:2005dc,Boos:2012zz,Chivukula:2021xod,Chivukula:2022tla}.

\newpage

\bibliographystyle{apsrev4-1.bst}

\bibliography{ref}{}

\end{document}